\documentclass[aps,prb,twocolumn,showpacs,amsmath]{revtex4-1}

\usepackage{braket}
\usepackage{amsmath}
\usepackage{amssymb}
\usepackage{graphicx}
\usepackage[colorlinks=true,citecolor=blue,linkcolor=red,linktocpage=true,pagebackref=false]{hyperref}

\usepackage[usenames,dvipsnames]{color}
\usepackage{varwidth}

\newcommand{\be}{\begin{eqnarray}}
\newcommand{\ee}{\end{eqnarray}}
\newcommand{\ben}{\begin{eqnarray*}}
\newcommand{\een}{\end{eqnarray*}}
\newcommand{\uu}{\ket{\uparrow\uparrow}}
\newcommand{\dd}{\ket{\downarrow \downarrow }}
\newcommand{\ud}{\ket{\uparrow \downarrow}}
\newcommand{\du}{\ket{\downarrow \uparrow}}

\usepackage{ulem}
\usepackage{soul}
\setstcolor{red}
\sethlcolor{yellow}

\begin{document}

\title{On-demand maximally entangled states \\
with a parity meter and continuous feedback}

\author{Clemens Meyer zu Rheda}
\affiliation{Dahlem Center for Complex Quantum Systems and Fachbereich Physik, Freie Universit\"at Berlin, 14195 Berlin, Germany}
\author{G\'eraldine Haack}
\altaffiliation[Present address: ] {D\'epartement de Physique Th\'eorique, Universit\'e de Gen\`eve, CH-1211 Gen\`eve 4, Switzerland\\}
\affiliation{Dahlem Center for Complex Quantum Systems and Fachbereich Physik, Freie Universit\"at Berlin, 14195 Berlin, Germany}
\author{Alessandro Romito}
\affiliation{Dahlem Center for Complex Quantum Systems and Fachbereich Physik, Freie Universit\"at Berlin, 14195 Berlin, Germany}

\date{\today}


\begin{abstract}
Generating on-demand maximally entangled states is one of the  corner stones for quantum information processing. Parity measurements can serve to create Bell states and have been implemented via an electronic Mach-Zehnder interferometer among others. However,  the entanglement generation is necessarily harmed by measurement induced dephasing processes in one of the two parity subspace. In this work, we propose two different schemes of continuous feedback for a parity measurement. They enable us to avoid both the measurement-induced dephasing process and the experimentally unavoidable dephasing, e.g. due to fluctuations of the gate voltages controlling the initialization of the qubits. We show that we can generate maximally entangled steady states in both parity subspaces. Importantly, the measurement scheme we propose is valid for implementation of parity measurements with feedback loops in various solid-state environments.
\end{abstract}

\maketitle

\section{Motivation}

The generation, the control and the read-out of pairs of entangled qubits serve as stepping stones towards the implementation of quantum information protocols. While the readout of individual qubits is typically associated with the irreversible destruction of the given coherent state, it has been shown that a joint measurement of two qubits can serve as an effective mechanism to generate entanglement between two measured qubits initially in a product state,\cite{Ruskov03, Engel05, Coish06, Ionicioiu07, Trauzettel06, Zilberberg08, Williams08, Kerckhoff09, Haack10, Lalumiere10, Tornberg10, Hofer13} and may be used for error correction schemes.\cite{Kitaev06, Fowler12} This measurement-based creation of entanglement is achieved by operating the detector as a parity meter. The corresponding observable in the computational basis of the qubits reads $\hat{P} = \hat{\sigma}_z^1 \otimes \hat{\sigma}_z^2$, where $\sigma_z^i$ are the Pauli matrices for the qubits $i=1,2$. The parity operator $\hat{P}$ has two eigenvalues, $\pm 
1$, corresponding to the even subspace spanned by the states $\{\uu,\dd\}$, and to the odd subspace spanned by $\{\ud, \du \}$. Measuring the parity causes the two-qubit state to collapse onto a superposition of even or odd states and, in particular onto the maximally entangled states or Bell states in an ideal situation, $\vert \Psi_e^{\pm} \rangle = (\uu \pm \dd)/\sqrt{2}$ and $\vert \Psi_o^{\pm} \rangle= (\ud \pm \du)/\sqrt{2}$.\\

\begin{figure}[hbt!]
\centering
 \includegraphics[width=8cm]{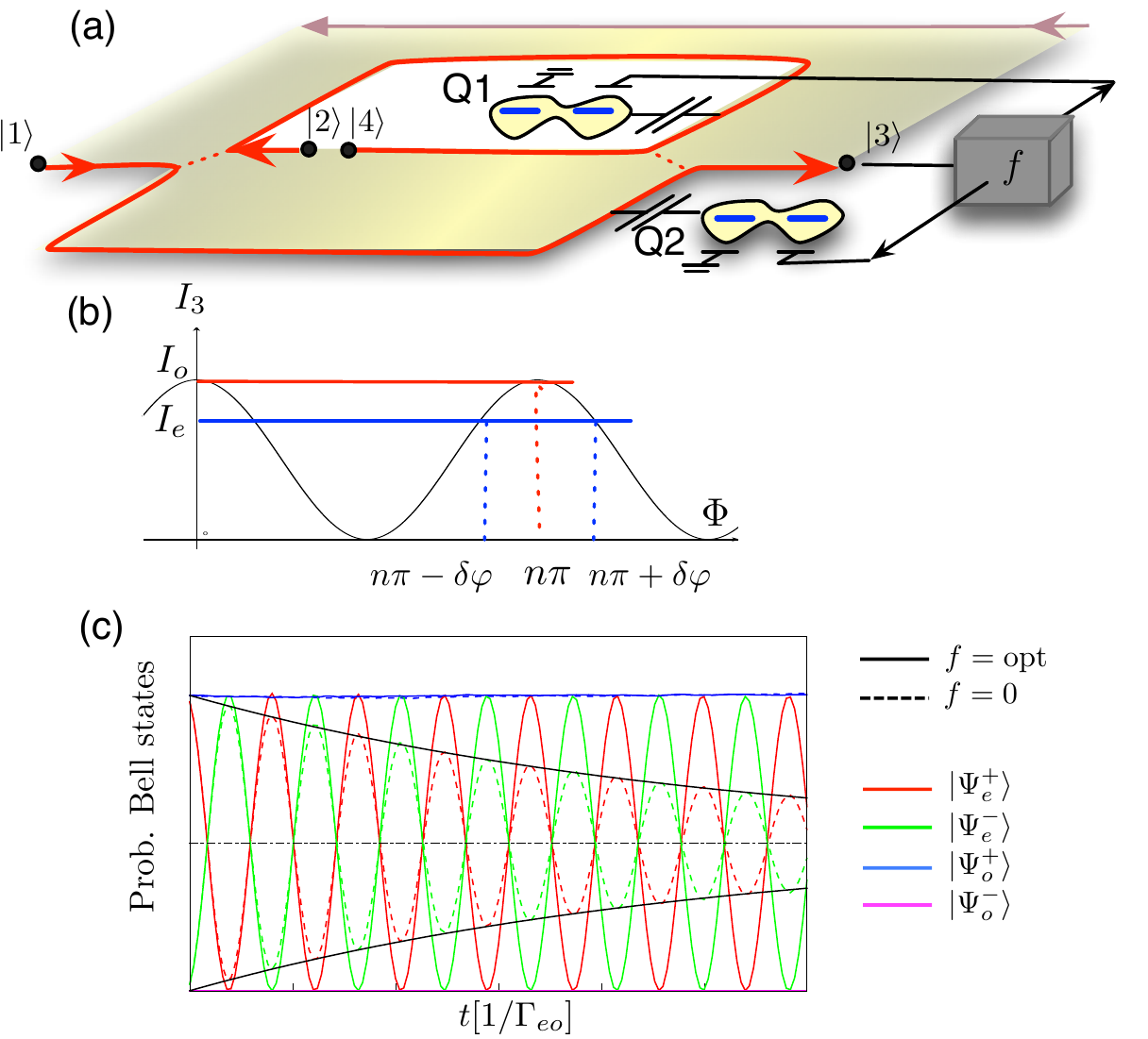}
\caption{(Color online). Parity measurement through a MZI and induced dephasing. (a) Sketch of the MZI operated as a parity meter for two charge qubits, $Q1$ and $Q2$, including the feedback circuit. The feedback is implemented via a gate voltage, which controls the bias energies of the qubits. Electrons are injected in lead $\ket{1}$ and the current is measured in lead $\ket{3}$. The input $\ket{2}$ and output $\ket{4}$ are not used in this proposal. (b) Signal at the detector's output. The Aharonov-Bohm flux $\Phi$ can be tuned to have indistinguishable signals for even and odd states, $I_e$ and $I_o$ respectively. (c) Probabilities for the two qubits to be in each of the four maximally entangled states as a function of time without feedback ($f=0$, dashed lines) and with optimal feedback ($f=\text{opt}$, solid lines).}
\label{fig1}
\end{figure}

Parity measurements have been proposed in various solid-state systems that serve as potential architectures for quantum computing. In circuit quantum electrodynamics (QED), generation of entanglement through a parity measurement has been very recently achieved in 3D circuit QED \cite{Riste13, Roch14} as well as in 2D Circuit QED \cite{Saira14}, an architecture suitable for surface coding. Also quantum transport-based measurements can equally well act as parity measurements: both the quantum point contact (QPC)\cite{Trauzettel06, Williams08} and the electronic Mach-Zehnder interferometer (MZI) \cite{Haack10} have been investigated as  parity meters for two double-quantum dots (DDs) charge qubits  as shown in Fig.~\ref{fig1}(a). By adjusting the coupling strengths between the qubits and the detector, and by tuning either the microwave resonator frequency or the magnetic flux threading the electronic MZI, only two values for the photon intensity or the electrical current corresponding to the two parity 
subspaces are detected.
\\

It has been shown theoretically~\cite{Lalumiere10, Tornberg10, Haack10, Haack12}, and confirmed in the experiments of Refs.~\onlinecite{Riste13, Roch14, Saira14}, that the two parity subspaces are not on an equal footing with respect to dephasing processes. While the two even states $\uu$ and $\dd$ give rise to the same measurement outcome within the parity measurement process, they differ by a phase factor which depends on the specific measurement outcome. Consequently, the different measurement backaction leads to a loss of phase coherence when averaging over an ensemble of realizations. This dephasing in the even subspace, which we characterize as intrinsic because it is induced by the unavoidable measurement backaction, does not affect the odd states. Therefore, entangled states made of a superposition of even states experience decoherence even in the absence of external noise. In this paper we address the possibility of including a feedback loop in the parity measurement scheme to  overcome this 
intrinsic dephasing.\\

The majority of previous theoretical works investigating the implementation of feedback loops focused on a Bayesian formalism. It consists in updating the density matrix elements according to a state estimation which depends on the full history of the state's evolution and of the measurement outcomes.\cite{Korotkov01, Ruskov05, Sarovar2005,  Jordan06} Experimentally, this feedback scheme has been successfully implemented to stabilize Rabi oscillations of a single-qubit subject to measurement-induced dephasing.\cite{Vijay12} This Bayesian procedure has to be compared to a Markovian or direct feedback, where the feedback is proportional to the measurement outcome.\cite{Wiseman94, Doherty00, Wang01} The Markovian feedback presents the strong advantage to be much less challenging numerically and experimentally, as it does not require real-time resolution of non-linear stochastic differential equations needed for the state estimation in the Bayesian approach. While in general the Bayesian approach to feedback leads to more 
accurate results, it has been shown that, in the case of QND measurements and in the absence of external noise sources, Markovian and Bayesian approaches coincide.\cite{Wiseman02} In this work, we show that the implementation of a Markovian feedback is sufficient to tackle the intrinsic dephasing present in the even subspace as discussed above.\\

To be concrete, we consider an architecture made of an electronic MZI \cite{Ji03, Neder06} coupled to two DDs charge qubits. This choice is motivated by the very good coherence properties of the MZI \cite{Roulleau08, Litvin08, Bieri09, Roulleau09, Huynh12} which make it an ideal candidate as a quantum detector.\cite{Chung05, Haack11} In Ref.~\onlinecite{Haack10}, one of the authors has derived the conditions under which it can be operated as an ideal parity meter. Measurement-induced entanglement has been shown in principle, but is not stable due to intrinsic and external sources of noise. The goal of this work is to go beyond these results, and to show that a Markovian feedback efficiently compensates the measurement-induced dephasing and that external sources of noise, unavoidable in realistic setups, can also be overcome by including a second feedback based on the Bayesian approach. With this, we demonstrate how to generate in a deterministic way, target maximally-entangled states that do not decay in time.\\

The paper is organized as follows. In Sect.~\ref{section2}, we present the microscopic derivation of the state at the output of the MZI after the transit of $N$ electrons, imposing that the detector acts as a parity meter. In Sect.~\ref{section3}, we analyze the average detector outcome and, based on the previous derivation of the output state,  derive the dephasing rates affecting the parity measurement. In particular, we discuss the dephasing rate $\Gamma_{eo}$ between the even and odd states which sets the time scale at which entanglement is generated, and we explain the origin of the dephasing rate $\Gamma_{ee}$ present within the even subspace. We introduce the concurrence conditioned on the parity outcome as a measure of the amount of entanglement generated in each parity class through the measurement process and advance a Markovian feedback scheme that tackles this intrinsic dephasing. To trigger future experiments, in Sect.~\ref{section5}, we expand our feedback mechanism to take into account the 
effect of additional sources of noise, especially when considering fluctuating tunnelling and bias energies in the DDs. To this end, we introduce a second joint measurement, whose outcome is used to implement a feedback based on the Bayesian approach. By combining the two feedback schemes, we achieve the deterministic generation of maximally-entangled states via a parity measurement, characterized by maximal concurrences in both parity subpsaces.

\section{Microscopic model} 
\label{section2}

The electronic MZI is made of a Corbino disk, built in the quantum Hall regime.\cite{Ji03, Neder06, Roulleau08, Litvin08, Bieri09, Roulleau09, Huynh12} The transport of electrons takes place along chiral edge states, which ensure a uni-directional transport. Two quantum point contacts (QPC) (left and right) act as beam splitters for the incoming electronic wave-packets. They are characterized by their reflection and transmission probabilities, $R_\mathrm{L,R}$ and $T_\mathrm{R,L}$, respectively.  The electrons are injected in lead $\ket{1}$, biased by an energy $eV$ compared to the other leads which are at the Fermi energy $E_F$. The DDs charge qubits are coupled capacitively to the arms of the MZI, as depicted in Fig.~\ref{fig1}.  Depending on the charge configuration of the DDs, each electron injected in the MZI acquires different phases in the upper ($u$) and lower ($d$) arms of the interferometer, $\varphi_u^{(ss')}$,$\varphi_d^{(ss')}$. Their expressions are derived from the interaction Hamiltonian 
between the charge in the DDs and the charge in the arms of the MZI.\cite{Haack10, Haack12, Pilgram02, Aleiner02}. The indices $s,s' \in \{ \uparrow, \downarrow \}$ span the DDs' computational basis. Starting with the qubits in a product state and decoupled from the detector described by the state $ \vert \Psi_\text{det} \rangle$, the initial state describing the whole system reads:
\be
 \label{eq:initial}
&&\Big(\alpha^{(\uparrow\uparrow)} \uu + \alpha^{(\uparrow\downarrow)} \ud + \alpha^{(\downarrow\uparrow)} \du + \alpha^{(\downarrow\downarrow)} \dd \Big) \otimes \vert \Psi_\text{det} \rangle\,, \nonumber \\
&& \quad
\ee
with $\vert\alpha^{(\uparrow\uparrow)}\vert^2 + \vert\alpha^{(\uparrow\downarrow)}\vert^2+\vert\alpha^{(\downarrow\uparrow)}\vert^2+\vert\alpha^{(\downarrow\downarrow)}\vert^2 =1$. 
After the passage of a single electron, the detector and the qubits become entangled:\cite{Ruskov03, Averin05, Jordan06, Williams08, Romito08, Haack10} 
\be
\label{eq:1e}
\vert \Psi_1 \rangle &=& \sum_{(ss')} \alpha^{(ss')} \Big[ C_3^{(ss')} \vert 3 \rangle+ C_4^{(ss')} \vert 4 \rangle \Big] \otimes \vert ss' \rangle \,,
\ee
where the states $\vert 3 \rangle$ and $\vert 4 \rangle$ denote the two output leads of the interferometer (see Fig.~\ref{fig1}) and the coefficients $C_3^{(ss')}$ and $C_4^{(ss')}$ in terms of the system's parameters are given by:
\be
C_3^{(ss')} \!&=&\! \sqrt{R_\mathrm{L} R_\mathrm{R}} e^{i2\pi \Phi/\Phi_0} e^{i\delta \varphi^{(ss')}_u}+ \sqrt{T_\mathrm{L} T_\mathrm{R}} e^{i\delta \varphi^{(ss')}_d}\nonumber \\
&=& \frac{e^{i2\pi \Phi/\Phi_0} e^{i\delta \varphi^{(ss')}_u} + e^{i\delta \varphi^{(ss')}_d}}{2} \,,\\
C_4^{(ss')} \!&=&\! i \big(\sqrt{R_\mathrm{L} T_\mathrm{R}} e^{i2\pi \Phi/\Phi_0} e^{i\delta \varphi^{(ss')}_u}- \sqrt{T_\mathrm{L} R_\mathrm{R}} e^{i\delta \varphi^{(ss')}_d} \big) \nonumber \\
&=& i \frac{e^{i2\pi \Phi/\Phi_0}e^{i\delta \varphi^{(ss')}_u} -e^{i\delta \varphi^{(ss')}_d}}{2}\,.
\ee
 The Aharonov-Bohm flux $2\pi \Phi/\Phi_0$, where $\Phi_0=h/e$ is the flux quantum, arises from the magnetic field threading the sample. The second equality for both coefficients is obtained by assuming symmetric QPCs, $R_\mathrm{L} = R_\mathrm{R} = 1/2$ in order to optimize the strength of the measured signal in agreement with the original experiments \cite{Ji03, Roulleau08, Bieri09}.\\

\textit{Parity detection--} To operate the electronic MZI as a parity meter, we have to impose that:
\be
\vert \langle \uparrow \downarrow \! \vert \Psi_1 \rangle \vert^2 &=&  \vert \langle \downarrow \uparrow \! \vert \Psi_1 \rangle \vert^2 = P_o \,. \label{eq:proba_parity1} \\
\vert \langle \uparrow \uparrow \! \vert \Psi_1 \rangle \vert^2 &=&  \vert \langle \downarrow \downarrow \! \vert \Psi_1 \rangle \vert^2 = P_e \label{eq:proba_parity2} \,.
\ee
where $P_o$ ($P_e$) is the probability to find the qubits in an odd (even) state. The two distinguishable currents at the output of the parity meter are then given by:
\be
I_o = \frac{ 2e^2V}{h} P_o \,, \quad I_e = \frac{2 e^2V}{h} P_e\,.
\ee                                          
Following Ref.~\onlinecite{Haack10}, Eqs~(\ref{eq:proba_parity1}) and (\ref{eq:proba_parity2}) are satisfied if the two parity assumptions are verified: $(P1): \, \, \delta \varphi_{u} = \delta \varphi_{d} \equiv \delta \varphi $, $(P2): \,\, 2\pi \Phi/\Phi_0 = 0 \,\, \text{mod} \pi$. Experimentally, the first condition can be fulfilled by controlling the capacitive coupling between  the arms of the MZI and the  DDs and the second one is directly implemented by tuning the magnetic field.\cite{Ji03, Roulleau08} The coupling is assumed to be weak such that:
\be
\Delta I = I_o - I_e \ll I_o, I_e\,. 
\ee
Consequently, the detector output noise $S_{\text{II}}$ is assumed to be independent of the qubits' state, $S_{\text{II}}= (S_{\text{II}}^o + S_{\text{II}}^e)/2$, with $S_{\text{II}}^{o/e} = 2 e^3 V/h \, T_{31}^{o/e}(1-T_{31}^{o/e})$ the detector shot noise associated with the transmission probability from lead $\ket{1}$ to lead $\ket{3}$ of the MZI for the qubits in the even and odd subspaces. The detector shot noise $S_{II}$ defines the measurement rate of the apparatus,\cite{Korotkov99} $\Gamma_\text{m} = \frac{(\Delta I)^2}{4 S_\text{II}}$. A quantum-limited ideal parity meter, i.e., a detector which detects the parity of the qubits as fast as it dephases them, is characterized by $\Gamma_\text{m} = \Gamma_{eo}$, where $\Gamma_{eo}$ describes the loss of phase coherence between the even and odd states. To get an explicit expression of $\Gamma_{eo}$, and of the dephasing rates within each subspace, $\Gamma_{ee}$ and $\Gamma_{oo}$, we expand the microscopic single-electron formulation of the output state $\vert \Psi_1 \rangle$ to 
the case of a large number $N$ of electrons and obtain an expression for $\ket{\Psi_N}$.\\ 

For this, we consider a time interval $\tau$, much larger than the time $h/eV$ between two consecutive electrons passing through the MZI, $\tau \gg h/eV$ (we recall that $eV$ sets the energy bias of the dc-source). During this time window $\tau$, $N \gg 1$ electrons are sent into the interferometer, independently from each other. Equation~(\ref{eq:initial}) then evolves to: 
\be
\label{eq:psi_N}
&&\ket{\Psi_N}=\sum_{ss'}\sum^N_n\genfrac(){0pt}{}{N}{n}C_3^{(ss')n}  C_4^{(ss')(N-n)} \ket{n,N-n}\ket{ss'}\,.\nonumber \\
&& \quad
\ee
Here, we have introduced the notation $\ket{n,N-n}$ to describe the state where $n$ electrons have reached the output lead $\ket{3}$ of the MZI, whereas $N-n$ have reached the output lead $\ket{4}$. The chirality of the edge states along which the electrons travel in the MZI ensures that the $N$ electrons injected into lead $\ket{1}$ exit either through lead $\ket{3}$ or lead $\ket{4}$.\\

 \begin{figure*}[t!]
\includegraphics[width=1.0\textwidth]{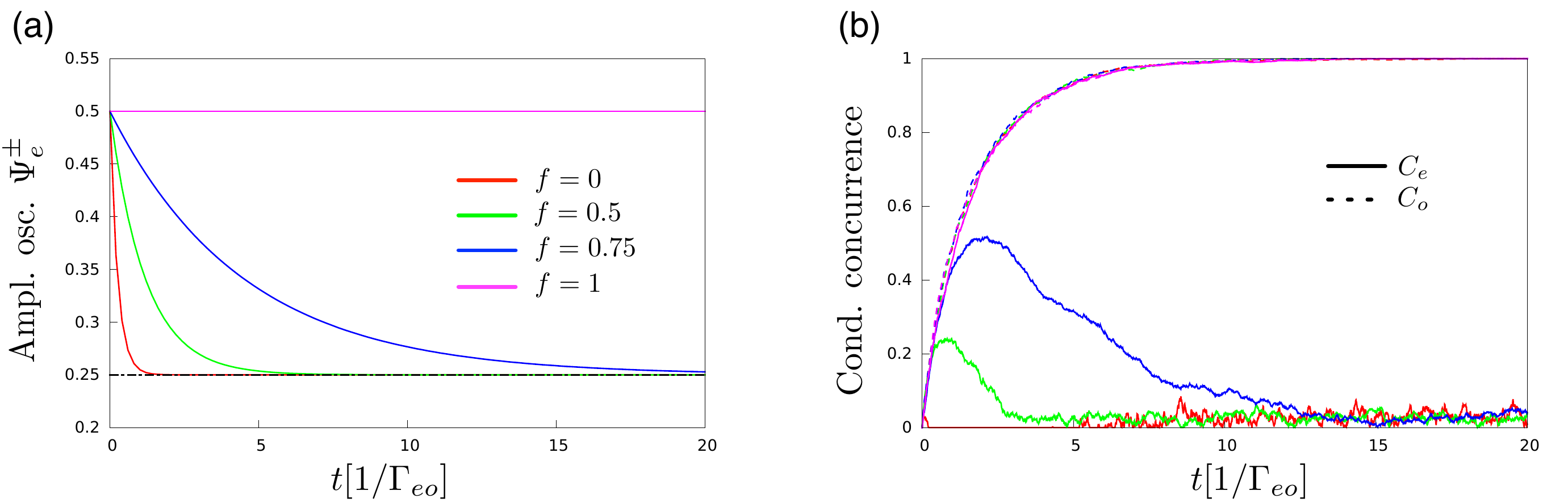}
\caption{(Color online). Generation of entanglement through the parity measurement with feedback. a) Amplitude of the oscillations between the populations of the even Bell states $\vert \Psi_e^{\pm} \rangle$ as a function of time for different values of the feedback parameter. For $f=1$, the amplitude does not decay as expected for the optimal feedback when considering a QND-measurement. b) Conditional concurrences plotted as a function of time in units of the measurement time set by the dephasing rate $\Gamma_{eo}$.  The optimal value for the feedback parameter $f=1$ leads to steady maximally-entangled states in both parity subspaces: $C_e= C_o$ do not decay. For comparison, conditional concurrences are shown for different values of non optimal feedback $f$: $f=0,0.5, 0.75$.}
\label{fig:feedback}
\end{figure*}

Though it is possible to inject a controlled train of single quasiparticles in a MZI,~\cite{Dublois2013} the most natural experimental configuration is that of an applied dc-bias. Indeed, a dc-source with a typical bias of $\sim 1 \mu \text{eV}$ will lead to the continuous emission of electrons with a typical coherence length $\sim 40 \mu \text{eV}$ (much longer than the coherence length of Levitons for instance, $\sim 1 \mu \text{eV}$, see Ref.~\onlinecite{Haack11} for details). Because the electronic MZI is of the order of few $\mu \text{eV}$, electrons sent by a dc-source can therefore be considered as interacting independently with the charges of the DQDs coupled to the arms of the interferometer.

 In the limit of $N \gg 1$ independent electrons, the central limit theorem applies and the binomial distribution in Eq.~(\ref{eq:psi_N}) tends to a Gaussian distribution. The output state $\vert \Psi_\text{out} \rangle$ in lead $\ket{3}$ (where the current is actually measured) is obtained by projecting the state $\vert \Psi_N \rangle$ onto the state $\vert n, N-n \rangle$. Expressing the numbers of electrons $n$ and $N$ in terms of the corresponding currents, $n = I_3 \tau/e$ and $N = I_1 \tau/e$ (the subindices refer to the leads, $e$ is the electrical charge and $\tau$ is the time interval at which the detector signal is registered), we 
obtain the output state (up to a normalization factor):
\be
\label{eq:Ne}
\ket{ \Psi_\text{out} } &\propto& \alpha^{(\uparrow\uparrow)} \frac{e^{-(I_3-I_e)^2/4D} }{\sqrt{4 \pi D}} e^{i I_1 \tau\delta \varphi} e^{i \pi (I_3-I_1)\tau} \uu \nonumber \\
&+& \alpha^{(\downarrow\downarrow)} \frac{e^{-(I_3-I_e)^2/4D}}{\sqrt{4 \pi D}} e^{iI_1 \tau \delta \varphi} e^{-i \pi (I_3-I_1)\tau}  \dd \nonumber \\
&+& \frac{e^{-(I_3-I_o)^2/4D}}{\sqrt{4 \pi D}} \big( \alpha^{(\uparrow\downarrow)} \ud + \alpha^{(\downarrow\uparrow)} e^{i 2 I_1 \tau \delta \varphi} \du\big)\,. \nonumber \\
&& \quad 
\ee 
The width $D$ of the two Gaussian distributions centered around the average values for the even and odd currents is set by the detector shot noise, $D = S_{II}/ \tau$. Equation~(\ref{eq:Ne}) constitutes one of the main results presented in this work and is in full agreement with previous theoretical expressions derived within a quantum Bayesian formalism.~\cite{Ruskov03, Ruskov05, Jordan06} Importantly, the parity measurement provides faithful information about the state of the system, as long as the width of the two Gaussian distributions is smaller than the signal $\Delta I = I_o - I_e$ to be resolved. The phase factor for the state $\du$ corresponds to a shift in energy of the qubits induced by the coupling to the detector (equivalent to an ac-Stark shift in Circuit QED \cite{Schuster05}).\\ 

The free evolution of the qubits during the time interval $\tau$  is generated by the Hamiltonian~$\hat{H}_\text{qb}$:
\be
\label{eq:freeevolution}
\hat{H}_{\text{qb}}= \sum_{i=1,2} \frac{\epsilon_i}{2} \hat{\sigma}_z^i + \Delta _i \hat{\sigma}_x^i \,,
\ee
 where $\epsilon_i$ and $\Delta _i$ are the energy bias and the tunneling energy of the qubits $i=1,2$ and the operators $\hat{\sigma}_z^i$ and $\hat{\sigma}_x^i$ are the Pauli matrices of the respective qubits. When the tunneling energy $\Delta$ is set to 0 and $\epsilon_i \equiv \epsilon$, the effect of the free evolution is merely adding a phase factor $e^{\mp i \epsilon \tau/\hbar}$ to the even states. Different bias energies do not change the dynamics in the even subspace and do not lead to additional dephasing in the odd subspace. As $\Delta=0$, this corresponds to a quantum non-demolition measurement (QND), $[\hat{H}_\text{qb}, \hat{P}]=0$. In the following, we derive the explicit expressions for the dephasing rates $\Gamma_{eo}, \Gamma_{ee}$ and $\Gamma_{oo}$ in the QND case. This shows that the parity measurement is of special interest as dephasing remains present, even when considering a QND measurement. It is the presence of dephasing in the even subspace that motivates the implementation of a feedback scheme.\\ 

\section{Intrinsic dephasing: Markovian feedback}
\label{section3}

\subsection{Dephasing processes}

The dephasing rates are obtained by averaging the ratios of the phase factors in Eq.~(\ref{eq:Ne}) over many realizations:
\be
\label{eq:gamma_oo}\Gamma_{oo} &=& -\frac{1}{\tau} \log \big\langle  e^{i 2 I_1 \delta \varphi \tau/e} \big\rangle = 0\,, \\
\label{eq:gamma_ee}\Gamma_{ee} &=& - \frac{1}{\tau} \log \big\langle e^{i 2\pi (I_3 - I_1) \tau/e} \big\rangle = \pi \frac{eV}{\hbar} \left( \frac{\delta \varphi^2}{4}\right),\\
\label{eq:gamma_eo}\Gamma_{eo} &=& - \frac{1}{\tau} \log \big\langle e^{i \pi (I_3 - I_1) \tau/e} \big\rangle = \frac{\Gamma_{ee}}{4}\,.
\ee
There is no dephasing present in the odd subspace as there are no fluctuating quantities. In fact electrons travelling through the interferometer acquire exactly the same phase on the upper and lower arms if the qubits are in the state $\ud$  or  $\du$. Contrarily, within the even subspace the dephasing rate $\Gamma_{ee}$  is finite. When the qubits are in an even configuration, the electrons acquire a different phase when traveling along the upper or the lower arms of the interferometer. The fluctuations in the phase factor of the states $\uu$ and $\dd$ is the hallmark of the quantum uncertainty concerning the path of the electrons when travelling through the MZI. The origin of $\Gamma_{ee}$ can also be understood from the detector's point of view: because the phase acquired by the electrons depends on the states $\uu$ and $\dd$ and is asymmetric between the upper and lower arms, the qubits acquire path information, which leads to dephasing.\\

Note that $\Gamma_{eo}$ sets the time scale at which entanglement is generated through the parity measurement process: it corresponds to $\Gamma_{eo} \equiv \Gamma_m = (\Delta I)^2/4 S_{II}$. Because the rate $\Gamma_{ee}$ is four times larger than $\Gamma_{eo}$, the parity measurement will never achieve the creation of significant entanglement in the even subspace if the corresponding dephasing is not compensated by a feedback loop.

\subsection{Markovian feedback mechanism}
As the effect of the fluctuating part of the intrinsic measurement backaction is proportional to the qubits' Hamiltonian the feedback can be simply implemented by adjusting the bias energy of the qubits as follows: 
\be
\label{eq:feedback}
\epsilon' = \epsilon + f (\pi (I_3-I_1) \frac{\hbar}{e})\,.
\ee
We recall that $I_1$ corresponds to the incoming current in lead $\ket{1}$, whereas $I_3$ is the measurement outcome. This procedure corresponds to a Markovian feedback, without any state estimation. Taking into account the last measurement outcome is sufficient here to stabilize the probabilities of generating the four maximally-entangled states as seen in Fig.~\ref{fig1}. We have assumed the feedback parameter to be the same for both qubits. This is justified as different $f$ would lead to extra dephasing in the odd subspace, which can be taken care by a more general feedback scheme discussed in Sect. \ref{section5}. In Fig.~\ref{fig:feedback} (a), we show the amplitude of the oscillations that the even Bell states under go. For the optimal value of the feedback parameter, $f=1$, this amplitude is maximal and does not decay in time.\\

The optimal value of $f$ can be derived from an analytical model based on the Langevin equation for the phase difference acquired by the qubits in the even subspace using Eqs.~(\ref{eq:Ne}) and (\ref{eq:feedback}):
\be
\label{eq:langevin}
\dot{\phi}(t) = \lim_{\tau \rightarrow 0} \frac{\phi(t+\tau)-\phi(t)}{\tau} = - \frac{\epsilon}{\hbar} + \pi \sqrt{D} \,( 1-f ) \,\xi(t) \,, \nonumber
\ee
where $\xi(t) = (I_3 -I_1)/\sqrt{D}$ is a  fluctuating variable characterised by a white noise: $\left \langle \xi(t) \xi(t')\right \rangle = 2 \delta(t-t')$. From the above equation, one deduces the corresponding Fokker-Planck equation for the probability distribution of the phase $\phi$:\cite{Risken96}
\be
\label{eq:fokker}
\frac{d}{dt} P(\phi, t) &=& (1-f)^2 \frac{\pi^2}{e^2} D \frac{d^2}{d \phi^2} P(\phi, t) \nonumber \\
&=& \frac{ (1-f)^2 \, 4 \, \Gamma_{eo}}{\tau} \frac{d^2}{d \phi^2} P(\phi, t) \nonumber \\
&\equiv& \frac{\Gamma_{ee}(f)}{\tau} \frac{d^2}{d \phi^2} P(\phi, t)\,.
\ee
Here $\Gamma_{ee}(f) /\tau$ corresponds to the diffusion coefficient of the phase $\phi$, given that the random variable depends on the output current $I_3$. From Eqs.~(\ref{eq:Ne}) and (\ref{eq:gamma_eo}), the expression of $D= 4 e^2 \Gamma_{eo}/\pi^2 \tau$ is derived. The optimisation of our feedback scheme corresponds to requiring that $\Gamma_{ee}(f)/\Gamma_{eo} = 4(1-f)^2 \equiv 0$. When the feedback parameter $f=0$, one recovers the dephasing rate $\Gamma_{ee}$ derived in Eq.~(\ref{eq:gamma_eo}) and for $f=1$, we reach the optimal situation where dephasing in the even subspace vanishes.\\

\textit{Numerical implementation--} To implement the continuous nature of the measurement and the feedback numerically, we consider two small consecutive time intervals $\Delta t$ large enough such that Eq.~(\ref{eq:Ne}) holds, but still short compared to the measurement time $1/\Gamma_{eo}$. At the end of the first interval, the state has evolved due to its bias energy $\epsilon$, and a current measurement result $I_3^i$ is drawn from the probability distribution pertaining to $I_3$. The state describing the system "qubits+detector" is then given by Eq.~(\ref{eq:Ne}). Consequently, the measurement result is used to adjust the bias energy $\epsilon$ to $\epsilon_{i+1} = \epsilon + f (\pi (I^{i}_3-I_1) \frac{\hbar}{e})$. During the second time interval, the state evolves under the Eq.~(\ref{eq:freeevolution}) with a bias energy $\epsilon_{i+1}$. Notice that for $\Delta t\rightarrow0$, measurement and evolution can be treated consecutively, also in the general case of $[\hat{H},\hat{P}]\ne0$.

\subsection{Deterministic generation of entanglement: \\ Conditional concurrences}

To describe quantitatively the amount of measurement-generated entanglement during the parity measurement with feedback, we introduce the conditional concurrence $C_{e/o}(\rho)$ defined as the entanglement concurrence post-selected on the measurement outcome. The concurrence is defined as:\cite{Wootters98}
\be
\label{eq:conc}
C(\rho) = \text{max}\left(0, \lambda_1 - \lambda_2 - \lambda_3 - \lambda_4\right)\,,
\ee
where $\lambda_i$ are eigenvalues of the matrix $R=\sqrt{\sqrt{\rho}\tilde{\rho}\sqrt{\rho}}$, $\rho = \sum_{s_1,s_1',s_2,s_2'} \alpha^{(s_1s_1',s_2s_2')} \vert s_1s_2' \rangle \langle s_2s_1' \vert$ denoting the two-qubits density matrix and $\tilde{\rho}$ its ``spin-flipped'' counterpart. Since the concurrence itself is a measure of entanglement for mixed states, we introduce the conditional concurrence defined as the concurrence post-selected on the measurement outcome (either $I_e$ or $I_o$).\cite{Saira14}. This allows us to further distinguish entanglement creation in the even and in the odd subspaces. Figure \ref{fig:feedback} displays the conditional concurrences $C_e$ and $C_o$ for different values of the feedback parameter $f$. As expected, changing the feedback parameter has no quantitative effect on the generation of entanglement in the odd subspace: $C_o=1$ and the dashed curves corresponding to different values of $f$ are barely 
distinguishable. In contrast, it is only for the 
optimal value of $f$ derived from the Fokker-Planck equation, Eq.~(\ref{eq:fokker}), that deterministic entanglement is created in the even subspace for long times compared to the measurement time. While these results have been derived considering the electronic MZI as parity meter for charge qubits, they remain valid for considering alternative solid-state architectures, for instance circuit QED setups.\\

For the sake of completeness, let us mention the case of finite tunnelling, $\Delta \neq 0$ in Eq.~(\ref{eq:freeevolution}). The operators $\hat{\sigma}_x^{1,2}$ tend to mix the two parity subspaces, which would render the parity measurement procedure ineffective to generate entanglement. However, this undesired effect can be avoided by controlling the strength of the parity measurement.  Indeed, for sufficiently strong measurements ($\Gamma_{eo}\gg\Delta/\hbar$), the projective nature of the parity measurement prevents the two parity subspaces from mixing. The concurrences $C_e$ and $C_o$ are then similar to the ones shown in Fig.~\ref{fig:feedback}.\\ 
 
We conclude this section by comparing our results with Ref.~\onlinecite{Riste13}, where a parity measurement in a 3D circuit QED architecture was implemented along with a \textit{projective} feedback to deterministically obtain a target Bell state. The \textit{projective} feedback consists in flipping one qubit when the measurement outcome indicates a state belonging to the even subspace. This single-qubit rotation results in states consisting of superposition of only odd states. However, dephasing leads to a maximum of created entanglement at some finite time. As this time is shorter than the measurement time (which sets the time at which a clear separation of parity subspaces is possible), the maximum fidelity obtained is $66\%$. We claim that our feedback mechanism is of particular relevance for this situation as for $f\rightarrow1$ the maximum entanglement is reached after an arbitrary large time, allowing for a distinct separation of parity subspaces and thus for a theoretical fidelity of $100\%$, i.e. a 
deterministic creation of target Bell states. To trigger future experiments in this direction, we consider in the next section external sources of noise, which are unavoidable in a realistic realization of our proposal.

\section{Additional sources of noise :\\ Bayesian feedback}
\label{section5}

\begin{figure}[t!]
\includegraphics[width=0.45\textwidth]{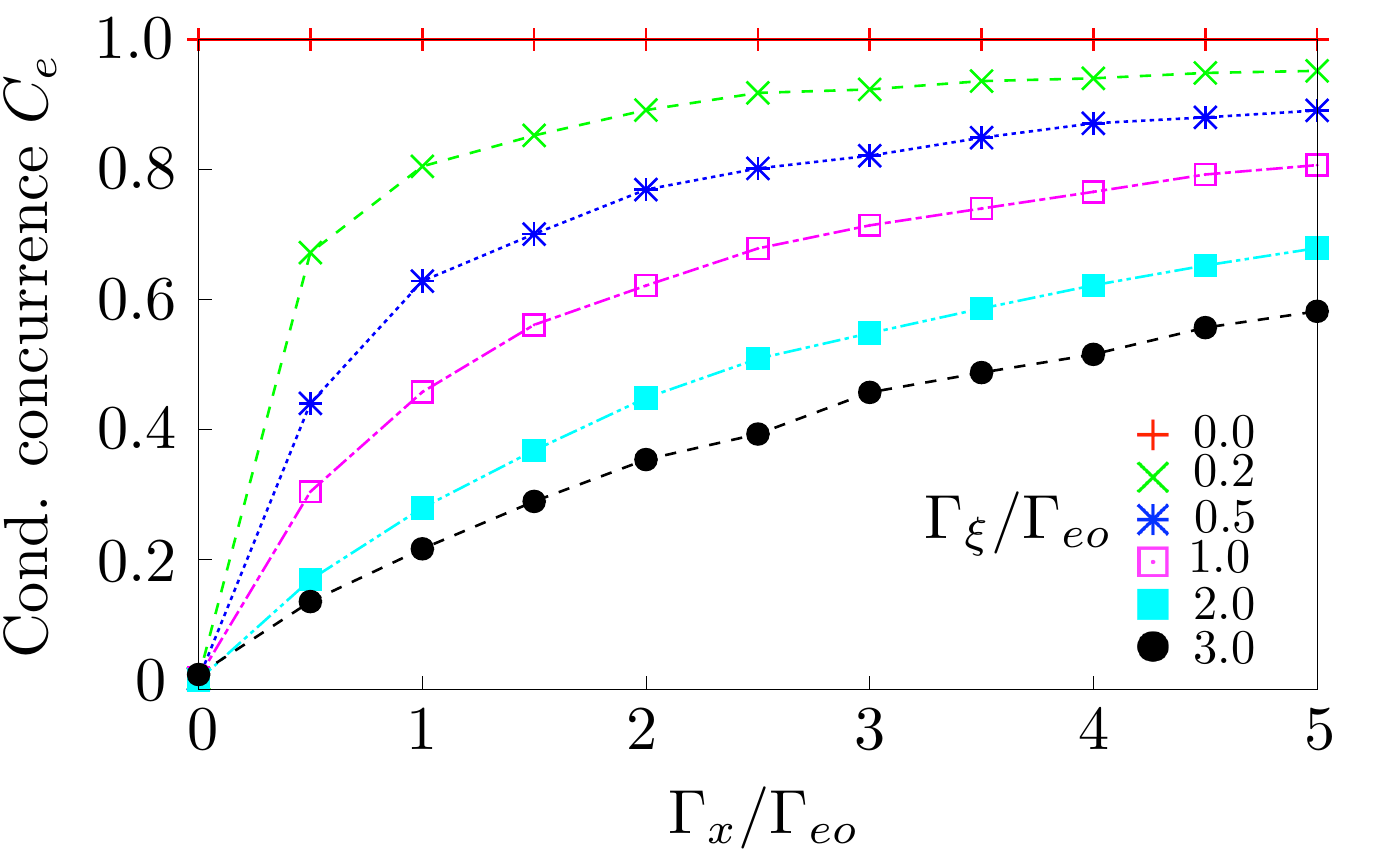}
\includegraphics[width=0.45\textwidth]{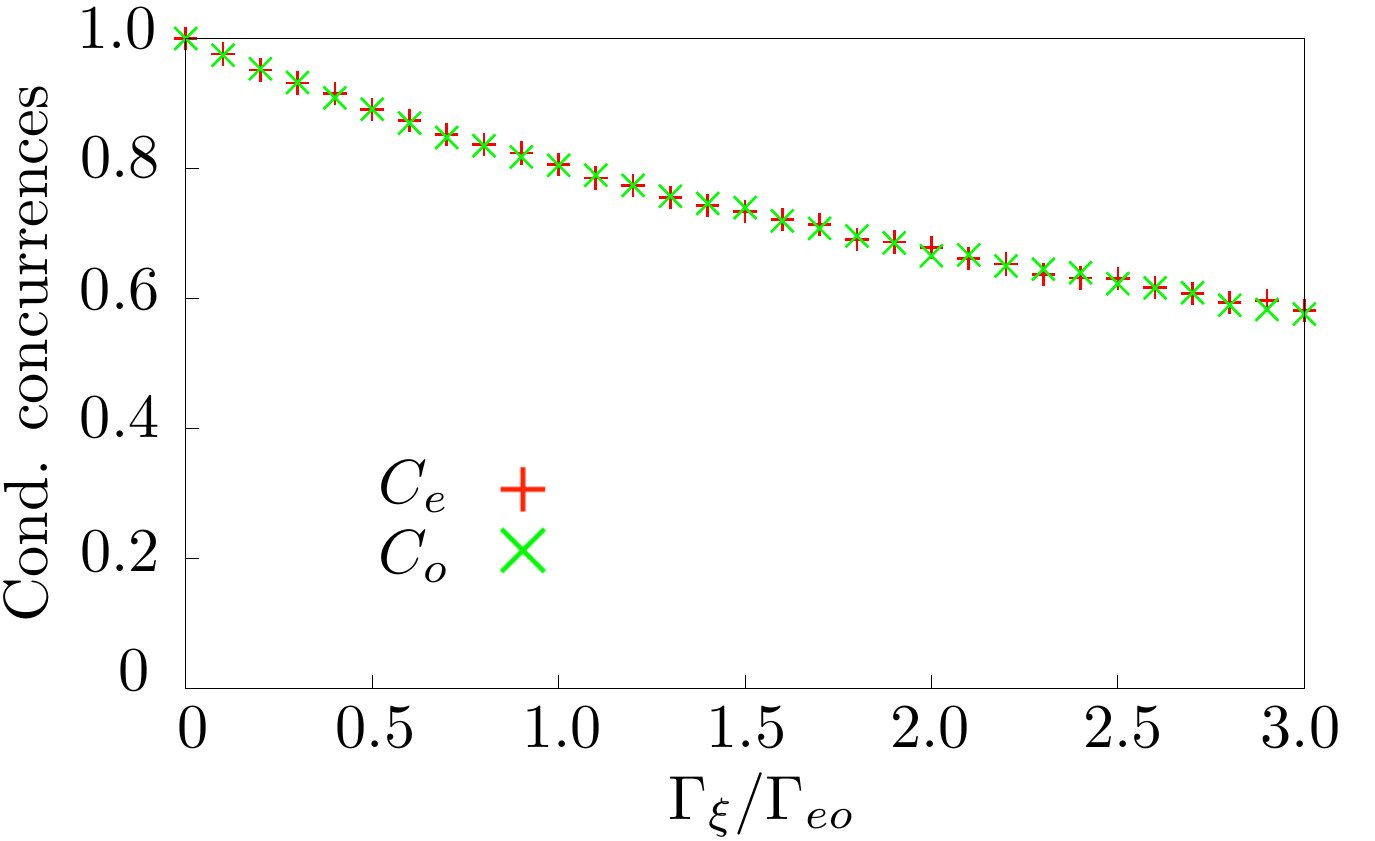}
\caption{(Color online). Deterministic generation of entangled states in the presence of external sources of noise. Upper panel: Conditional concurrence $C_e$ as a function of the measurement strength $\Gamma_x$ normalized by $\Gamma_{eo}$ for different noise strengths $\Gamma_\xi/\Gamma_{eo}$. In the absence of external noise, the Markovian feedback that compensates the intrinsic dephasing is enough to produce Bell states which do not decohere in time: $C_e=1$ (red crosses). For non-zero values of noise, $C_e$ reaches a saturated value which decreases when the noise strength becomes large. Lower panel: Conditional concurrence $C_o$ follows the same evolution as $C_e$: here the behaviour of $C_o$ as a function of the noise strength $\Gamma_\xi/\Gamma_{eo}$ is plotted.}\label{fig:finalentanglement}
\end{figure}

Deterministic generation of given Bell states requires an initial initialization of the qubits.  Experimentally, this requires  tuning of gate voltages, which induces  fluctuations in the energy and tunnelling energies of the qubits, and therefore leads to additional dephasing. To mimic this situation, we assume that the noise sources of the qubits are statistically uncorrelated, so that their Hamiltonian now reads:
\be
\label{eq:noise}
\hat{H}_{qb} = \sum_{i=1,2} \left( \frac{\epsilon_i}{2} + \xi_i(t) \right) \hat{\sigma}_z^i\,,
\ee
where the fluctuating terms $\xi_i, i=1,2$ are separately drawn from a white noise distribution with a width $D_{\xi}$. This leads to an extra dephasing within each subspace, characterized by the rate $\Gamma_\xi$.  
Treating the additional dephasing along with the intrinsic dephasing requires that the detector output is a continuous variable on such time scale, \textit{i.e.} $\Delta t$ has to remains the shortest time-scale of the problem. In particular $\Gamma_\xi \, \Delta t \ll1$. With a decoherence time $1/\Gamma_\xi \sim10^{-8}\textrm{ seconds}$ for charge qubits~\cite{Petta2010} and a typical impinging rate $eV/\hbar \sim10^{10} \textrm{ seconds}^{-1}$ in MZI interferometers~\cite{Neder06, Roulleau08}, the condition can be met in current and future experiments. 
We assume the general case of non-symmetric bias energies $\epsilon_i$, but we do not consider noise in the tunnelling. This is justified by realizing that the consequence of a noisy tunnelling can be split into two distinct effects, which can be easily compensated. First, it induces a dephasing between the parity subspaces that can be suppressed by a sufficiently strong parity measurement as discussed in the previous section. Secondly, it causes a fluctuating phase within the parity subspaces, which can be  reabsorbed in the fluctuating part of the Hamiltonian Eq.~(\ref{eq:noise}).\\

To the contrary, the presence of bias energy fluctuations results in a dephasing within both subspaces so that, for any finite $D_\xi$, the built-up entanglement reaches a maximum at some finite time and decreases afterwards down to zero.\cite{Ruskov03, Ruskov05, Riste13} These dephasing processes can not be compensated by a Markovian feedback based on the parity measurement result. Indeed, because $[ \hat{H}_{qb}, \hat{P}]=0$, the parity measurement does not infer any information about the loss of phase coherence and its outcome is then of no use for a feedback procedure aiming at compensating external sources of noise. We therefore introduce a new measurement operator,
\be
\label{eq:secmeas}
\hat{P}_x = \hat{\sigma}_x^1 \otimes \hat{\sigma}_x^2\,.
\ee 
This operator has the advantage to distinguish between states within one parity subspace while commuting with the parity operator $\hat{P}$.
Thus its measurement allows us to infer information about the relative phase between states in the even or in the odd subspace. Experimentally, this joint measurement can be implemented by pulsed measurements, combining single-qubit operations ($\pi/2$-rotations) with a parity measurement. \cite{Saira14, Leo_private} \\

To obtain more accurate results with this feedback loop based on $\hat{P}_x$, we consider here a Bayesian approach.\cite{Ruskov03, Vijay12, Roch14} Because $[ \hat{P}_{x}, \hat{P}]=0$, the measurement of $\hat{P}_x$ does not mix the parity subspaces. Therefore, the dynamics of the system, along with the joint measurement of $\hat{P}$, and $\hat{P}_x$ leads, at long times, to a decay of the density matrix elements connecting states of different parity. The evolution of the state at long time, and the effect of feedback, can then be characterized by the two phases expressed in terms of the real and imaginary parts of the density matrix' elements $\rho_{ss',ss'}$ written in the computational basis of the two qubits spanned by the indices $ss'$:
\be
\label{eq:phase_Bayes}
 \phi_e = \arctan{\frac{\text{Re}\{\rho_{\uparrow\uparrow,\downarrow\downarrow}\}}{\text{Im}\{\rho_{\uparrow\uparrow,\downarrow\downarrow}\}}},\quad \phi_o = \arctan{\frac{\text{Re}\{\rho_{\uparrow\downarrow,\downarrow\uparrow}\}}{\text{Im}\{\rho_{\uparrow\downarrow,\downarrow\uparrow}\}}}\,.
\ee
Compared to the free evolution of the two-qubit system, phase differences of
 \be
\label{eq:phase_diff_Bayes}
\Delta \phi_e = \left(\epsilon_1 + \epsilon_2\right) \tau - \phi_e,\quad \Delta \phi_o = \left(\epsilon_1 - \epsilon_2\right) \tau - \phi_o\,
\ee
have to be accounted for by the feedback mechanism within each subspace. Intrinsic dephasing is therefore compensated by the Markovian feedback and external sources of noise by the Bayesian one. Accounting for these two procedures, the bias energies $\epsilon_i$ of the qubits are updated as following:
\be
\epsilon_1 &\rightarrow& \epsilon_1' + f_x  \frac{\Delta\phi_e+\Delta\phi_o}{2} \\
\epsilon_2 &\rightarrow& \epsilon_2' + f_x \frac{\Delta\phi_e-\Delta\phi_o}{2} \,,
\ee
where the energies $ \epsilon_1'$ and $ \epsilon_2'$ are given by Eq.~(\ref{eq:feedback}). Though the experimental implementation of Bayesian feedback is more challenging than the direct feedback, this procedure remains realistic as the feedback is limited to one control parameter.\cite{Ruskov05, Jordan06, Vijay12} \\

Figure~\ref{fig:finalentanglement} shows the resulting conditional concurrences as a function of the strength of the second measurement $\hat{P}_x$. Here we model the detector noise by its variance $D_x$, the induced dephasing by $\Gamma_\xi$ (see Eq.~(\ref{eq:noise}) and below) and the interaction between the qubits' operators and the observable by a coupling parameter $\lambda_x$. These parameters determine the measurement strength of $\hat{P}_x$, $\Gamma_x \equiv \lambda_x^2 / D_x$.\cite{Clerk10} The concurrences are plotted as a function of $\Gamma_x$ normalized by the measurement rate $\Gamma_{eo}$, for different normalized values of the noise $\Gamma_\xi/\Gamma_{eo}$. \\

For no external noise, $\Gamma_\xi/\Gamma_{eo}=0$, the second feedback loop through the measurement of $\hat{P_x}$ is superfluous, the conditional concurrence $C_e$ remains maximal for all the values of $\Gamma_x/\Gamma_{eo}$ (red crosses). This corresponds to the results presented in the precedent section on the Markovian feedback. When noise in the bias energies is taken into account, the second measurement $\hat{P}_x$ is necessary to infer information about the loss of phase coherence. The conditional concurrence $C_e$ saturates when increasing the measurement strength $\Gamma_x$. This is consistent with previous theoretical works on a single two-level system.\cite{Ruskov05} The curves in Fig.~\ref{fig:finalentanglement} show that entanglement is generated and stabilized, even in the presence of external sources of noise, when we combine our Markovian and Bayesian procedures. For completeness, the lower panel shows that conditional concurrences in the odd and even subspaces behave uniformly.

\section{Conclusions}

In this work we have analyzed different feedback schemes for parity measurement of two qubits and considered explicitly the parity measurement of two DDs by a properly tuned MZI. We have first introduced a microscopic description of the measurement process which allowed us to quantify the measurement backaction in the even-parity subspace, where dephasing is dominant. This enabled us to implement a Markovian feedback, numerically much less demanding that the Bayesian one. Introducing the concurrence as an appropriate measure of entanglement, we showed that one can optimally tune the feedback strength to overcome the decay of entanglement. However, this feedback cannot overcome external sources of noise, unrelated to the measurement. To tackle these extra-dephasing processes, we have analyzed a more elaborate feedback scheme requiring the measurement of an additional joint operator. Its measurement outcomes are then used for a state estimation of the qubits, allowing for the implementation of a second 
feedback 
loop based on the Bayesian approach. The combination of the two feedback loops leads to stable entangled states, also in the presence of external noise sources. Though the analysis in our work is based on a Mach-Zehnder interferometer detection of double quantum dots, our protocols, the general formalism, and the results are generally valid for any implementation of a parity measurement in various solid-state architectures.

\section*{Acknowledgements}

We would like to thank motivating and insightful discussions with L. DiCarlo, A. N. Jordan, I. Neder and N. Roch. We acknowledge the financial support of DFG under Grant No. RO 4710/1-1. G.H. also acknowledges support from the Alexander von Humboldt Foundation and from the Swiss NCCR on Quantum Science and Technology (QSIT).

\end{document}